\documentclass[twocolumn,twoside,slac_two]{revtex4}
\usepackage{graphicx}
\usepackage{fancyhdr}
\newcommand{\Ms}{M_{\odot}}

\newcommand{\gauss}{\mathrm{G}}
\newcommand{\erg}{\mathrm{erg}}
\newcommand{\psec}{\mathrm{/s}}
\newcommand{\gpcmc}{\mathrm{g/cm^3}}

\newcommand{\ergps}{\rm ergs \,\, {s}^{-1}}

\begin{document}

\title{Long-Term Evolution of Collapsars: \\Mechanism of Outflow Production}

%

\author{Seiji Harikae}
\affiliation{  University of Tokyo}
\affiliation{  National Astronomical Observatory of Japan}
\author{Tomoya Takiwaki}
\affiliation{  National Astronomical Observatory of Japan}
\author{Kei Kotake}
\affiliation{  National Astronomical Observatory of Japan}
\affiliation{  Center for Computational Astrophysics, Japan}

\begin{abstract}
We present our numerical results of 
 two-dimensional hydrodynamic (HD) simulations and magnetohydrodynamic (MHD) simulations 
of the collapse of rotating massive stars 
in light of the collapsar model of gamma-ray bursts (GRBs).
Pushed by recent evolution calculations of
  GRB progenitors, we focus on lower angular momentum of the central 
core than the ones taken mostly in previous studies.
  By performing special relativistic simulations including 
both realistic equation of state and neutrino cooling, we follow a long-term evolution of the slowly rotating collapsars up to $\sim$ 10 s,
accompanied by the formation of jets and accretion disks. 
We find such outfows can be launched both by MHD process and neutrino process. 
We investigeate the properties of these jets whether it can become GRBs or 
remains primary weak outflow. 
\end{abstract}

\maketitle

\thispagestyle{fancy}


\section{Introduction}
Gamma-ray bursts (GRBs) are one of the most energetic phenomena
 in the universe. 
Pushed by recent observations, the so-called collapsar 
 has received quite some interest for the central engines of the 
 long GRBs \cite{woos93,pacz98,macf99}. 
Here we focus on the outflow formation in collapsars. 
There are two promising senario to launch outflow from collapsar, 
MHD process and/or neutrino process, 
which has also been extensively investigated thus 
far (e.g., \cite{macf99,prog03c,mizu06,fuji06,mcki07b,birk08,naga07,naga09}). 

A general outcome of the MHD studies among them 
is that if the central progenitor cores 
 have significant angular momentum ($\approx 10^{17}$ $\rm {cm}^2$/s) 
with strong magnetic fields ($\sim 10^{11}$ G), 
magneto-driven jets can be launched strong 
enough to expel the matter along the rotational axis within several seconds after the 
onset of collapse. 
  It should be noted that the angular momentum of those GRB progenitors 
($\approx 10^{16}$ $\rm {cm}^2$/s), albeit not a final answer due to 
 much uncertainty, is relatively smaller than those assumed in most of the 
previous collapsar simulations. 
These situations motivate us to focus on slower rotation of the central core
in the context of collapsar models \cite{hari09a}. 
As for the initial magnetic fields, we choose to explore relatively smaller 
fields ($\le 10^{10}$ G), which has been less investigated so far.
 Paying particular attention to the smaller angular momentum, it takes much 
longer time to amplify the magnetic fields large enough to launch the MHD jets than 
 previously estimated. By performing long-term evolution over $\sim$ 10 s, 
 we aim to clarify how the properties and the mechanism of the MHD jets could 
change with the initial angular momentum and the initial magnetic fields. 

In terms of the neutrino-driven outflows, it still remains preliminary ones, 
such as evaluation in the post processing manner \cite{birk08}.  
We have developed code to 
calculate neutrino heating via neutrino-antineutrino pair annihilation 
with ray-tracing method \cite{hari09b}. We show some result of the test 
calculation and application to the collapsar model.

\section{Magneto-Driven Jet from Collapsars}
\subsection{Numerical Methods and Initial Model}
\label{sec:num}
The results presented here are calculated by the MHD code 
in special relativity developed by \cite{hari09a,taki09},
 including both realistic equation of state
 and neutrino cooling. 
As for the initial profiles of the collapsing star, 
we employ the spherical data set of density, temperature, internal energy, 
and electron fraction in model 35OC in \cite{woos06}. 
We add cylindrical rotation and dipole magnetic field profiles with 
vanishing toroidal magnetic field in a parametric manner. 
We compute 15 models changing the initial rotation and 
the strength of magnetic fields (See \cite{hari09a} for details). 
Model name is given as 'BXJY' representing the model with 
$B=10^{X} \gauss$ and $j=Yj_{\rm lso}$, where 
$j$ is the specific angular momentum and $j_{\rm lso}$ is the specific angular momentum for the last stable 
angular orbit. 

\subsection{Result}
\label{sec:results}
Computing 15 models in a longer time stretch than ever among previous collapsar models,
we observe a wide variety of the dynamics changing drastically with time.
Here we summarize main results of this study. 

To extract the general features furthermore among the models, we focus on 
the properties of MHD outflows and neutrino luminosities.
It is noted that both of them is helpful to understand the energy sources 
 for powering the GRBs namely via magnetic and/or neutrino-heating 
mechanisms.  
\begin{table}[tb]
\begin{center}
\begin{tabular}[c]{|c|c|c|c|c|}
\hline
 Model& B10& B9& B8\\
\hline
J1.5& 
\shortstack{ $\bigcirc$ (TYPE II) \\ $1.1 \times 10^{52}$ erg/s \\ 3.4 s}& 
\shortstack{ $\bigcirc$ (TYPE I)  \\ $1.6 \times 10^{52}$ erg/s \\ 9.2 s}& 
\shortstack{ $\times$  \\ $1.8 \times 10^{52}$ erg/s \\ 4.3 s}
\\ \hline
J2.0& 
\shortstack{ $\bigcirc$ (TYPE II) \\ $4.5 \times 10^{51}$ erg/s \\ 5.8 s}& 
\shortstack{ $\bigcirc$ (TYPE I)  \\  $5.1 \times 10^{51}$ erg/s\\ 5.3 s}& 
\shortstack{ $\times$  \\ $8.5 \times 10^{51}$ erg/s \\ 6.0 s}
\\ \hline
J2.5& 
\shortstack{ $\bigcirc$ (TYPE I)  \\ $1.6 \times 10^{50}$ erg/s\\ 7.7 s}& 
\shortstack{ $\times$ \\ $1.4 \times 10^{50}$ erg/s \\ 12 s}& 
\shortstack{ $\times$  \\ $1.7 \times 10^{50}$ erg/s \\ 10 s}
\\ \hline
J3.0& 
\shortstack{ $\bigcirc$ (TYPE I)  \\ $9.0 \times 10^{49}$ erg/s \\ 9.3 s}&
\shortstack{ $\times$ \\ $2.5 \times 10^{49}$ erg/s \\ 14 s}&
\shortstack{ $\times$  \\ $4.5 \times 10^{49}$ erg/s  \\ 12 s}
\\ \hline
\end{tabular}
\caption{Properties of MHD jets and neutrino luminosities.
Contents of each cell is, whether the MHD jets are formed
(yes or no indicated by $\bigcirc$ or $\times$) with the different formation 
 mechanisms indicated by TYPE I or TYPE II (top),  the neutrino luminosity (middle)
 estimated at the epoch (bottom) when the accretion disks become almost stationary. 
}\label{Table:lumi_outflow}
\end{center}
\end{table}

Top column of each cell of Table \ref{Table:lumi_outflow} 
indicates whether the MHD jets are formed 
(yes or no indicated by $\bigcirc$ or $\times$).
TYPE I or II indicates the difference of the 
formation process of the MHD jets (see below). 
The quantities of the middle column show the neutrino luminosity 
(sum of all the neutrino species $\nu_e$, $\bar{\nu}_e$, and $\nu_{X}$) 
estimated at the epoch when the accretion disks becomes 
 stationary(bottom) (e.g., typically  $\sim$ 4 sec).
 We find that the neutrino luminosities become higher for slower rotation models.
 This is because the accretion disks can attain higher temperatures due to 
the gravitational compression. It is interesting to note that the luminosities
  tend to become smaller for strongly magnetized models with relatively 
 smaller angular momentum (J $\leq$ J2.5). This is mainly 
 because the gravitational compression is hindered by the magnetic forces confined 
 in the disks.
 
 \subsubsection{amplification of magnetic field and formation of magnetic wind}
Since the initial models investigated here are assumed to have only the poloidal
fields, 
 the key ingredients for amplifying the toroidal fields 
are the compression of the poloidal fields 
and the efficient wrapping 
of them via differential rotations. 
In addition, MRI should also play an important role, 
whose wavelength of the fastest growing mode is 
given by 
$\lambda \sim 5 
\Bigl( \frac{300\psec}{\Omega}	  \Bigl) 
\Bigl( \frac{B}{10^{12}\gauss}	  \Bigl) 
\Bigl( \frac{10^{10}\gpcmc}{\rho} \Bigl)^{1/2} \rm{km}$.  
Here, putting the typical physical values of the disk,
our numerical grids are insufficient to capture MRI at earlier
phase when the magnetic field is weak, but can handle it in
the later phase when the magnetic field gets stronger. In this
sense, the discussion below should give a lower bound for the field
amplification.

\begin{figure}[tb]
\begin{center}
\includegraphics[width = 80mm]{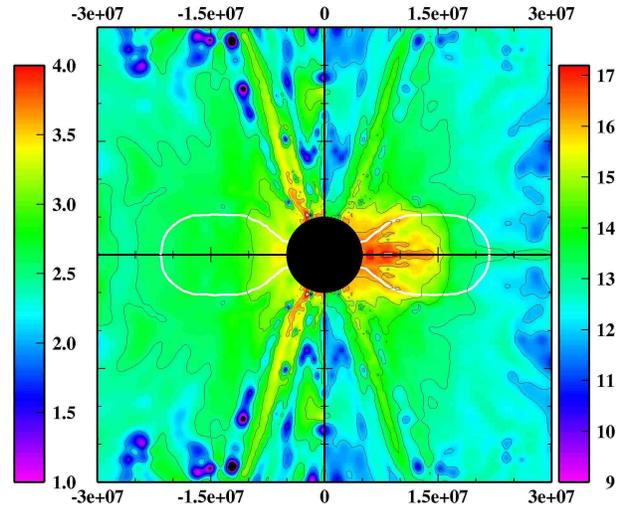}
\caption{Logarithmic contour of 
 differential rotation (:$Xd\Omega/dX$) (left) and amplification rate 
of toroidal magnetic field (:$d B_{\phi}/dt$) (right).}
\label{B9J1.5domegadBphi}
\end{center}
\end{figure}
Figure \ref{B9J1.5domegadBphi} shows the differential rotation and amplification rates of 
toroidal magnetic field of model B9J1.5
 at 7.99 s, just before the launch of the MHD outflows from the accretion disk.
The density takes its maximum value at around 120 km in the equatorial plane
 (left panel), in which the poloidal fields are strong because the higher 
compression is achieved (right panel). 
The white solid line in the right panel indicates
 the surface positions of the accretion disk. So the amplifications of the 
poloidal fields occur most efficiently inside the accretion disk.
It is noted here that the disk is gravitationally stable because 
the adiabatic index ($\gamma$) inside the disk becomes greater than $4/3$ due 
to the contribution of the non-relativistic nucleon ($\gamma = 5/3$) 
photodissociated from the iron nuclei. 
Figure \ref{B9J1.5domegadBphi} shows that the 
 amplification rates of toroidal magnetic field are highest also inside the disk (right), because the 
 degree of the differential rotation is large there (left). 
In previous collapsar simulations assuming much larger angular momentum initially, 
it seems to be widely agreed that the differential rotation is a primary agent to 
 amplify the toroidal fields. 
On the other hand, our results show that for 
long-term evolution of relatively slow rotation models, 
the amplification of the poloidal fields by compression 
 is preconditioned for the amplification of the toroidal fields. 

\subsubsection{two types of megneto-driven outflow}
\begin{figure}[tbd]
\begin{center}
\includegraphics[width=80mm]{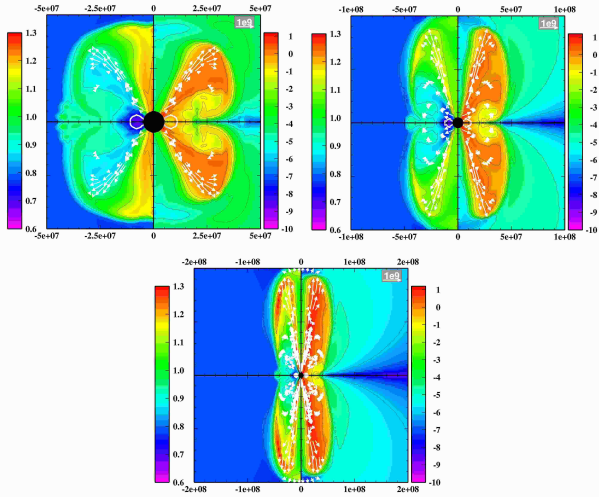}
\caption{Evolution of MHD jets launched from the accretion disk (Type II jets).
In each panel, logarithmic contour of 
 entropy (left side) and the inverse of plasma beta 
(e.g., $\beta_{\rm mag}^{-1} \equiv 
B^2/8\pi/p$) (right side) are shown at 1.43 s (top left), 
1.87 s (top right), and 2.11 s (bottom) for model B10J1.5.
Note the difference of the length scales among panels. }
\label{B10J1.5spmag}
\end{center}
\end{figure}
\begin{figure}[tbd]
\begin{center}
\includegraphics[width=80mm]{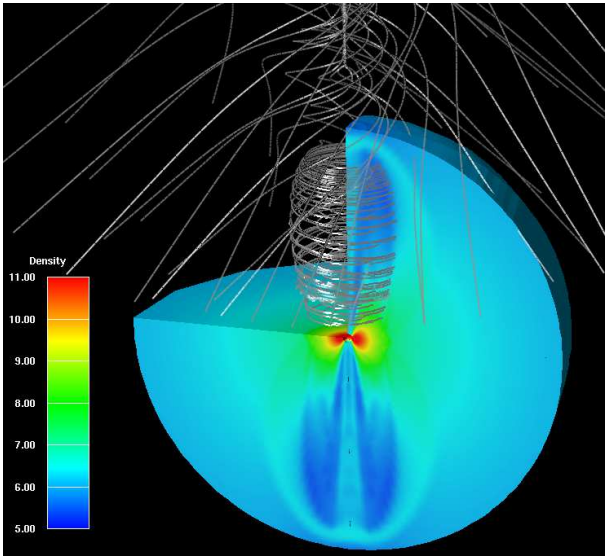}
\caption{Three dimensional plot of density with the magnetic field lines
 (silver line) for B10J1.5 model near at the moment of the shock break-out (2.11 s). 
Color contour on the two-dimensional slice represents the logarithmic density. }
\label{B10J1.5dB_3D}
\end{center}
\end{figure}
\begin{figure}[tbd]
\begin{center}
\includegraphics[width=80mm]{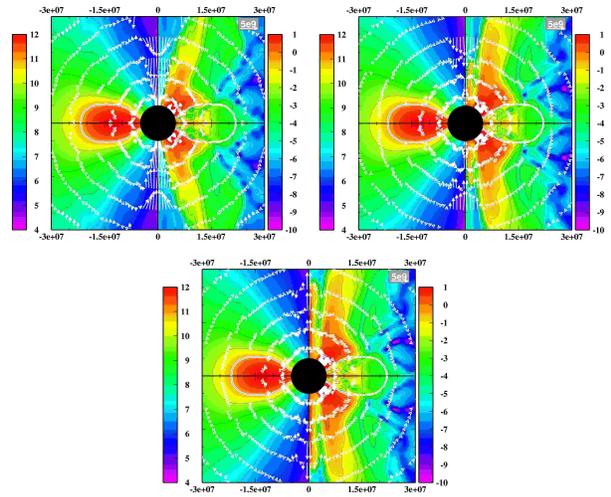}
\caption{Same as Figure \ref{B10J1.5spmag} but for model B9J1.5
 at 8.115 s (top left), 8.148 s (top right), and 8.154 s (bottom), 
 showing the moment of the formations of jets in type I.}
\label{B9J1.5dpmag}
\end{center}
\end{figure}
Figure \ref{B10J1.5spmag} shows the evolution of the MHD outflow for model B10J1.5,
 from its initiation near from the inner edge of the accretion disk (top left),
 propagation along the polar axis (top right), 
till they come out of the iron core (bottom). Note the difference of the length scales 
 in each panel. Among the computed models, this model has the strongest 
initial magnetic fields with smallest angular momentum (e.g., Table \ref{Table:lumi_outflow}). 
 The toroidal fields can be much stronger than for other models, by the 
 enhanced compression of the poloidal fields inside the disk. 
As a result, the MHD outflow is so strong that they do not 
 stall, once they are launched. In fact, the outflow is shown to be kept 
magnetically-dominated (inverse of the plasma beta greater than 1, right-hand side 
 in Figure \ref{B10J1.5spmag}) till the shock break-out.
 As the outflow propagates rather further from the center (top right),
 the outflow begins to be collimated due to the magnetic hoop stresses, 
 and keep their shape till the shock-breakout (Figure \ref{B10J1.5dB_3D}).
 
 We move to discuss the jets in type I, by taking model B9J1.5 as 
an example. In this case, the magnetic outflow from the accretion disk 
 is not as strong as model B10J1.5, and stalls at first in the iron core 
(see butterfly-like regions colored by red in the top left panel (right-hand side) 
 in Figure \ref{B9J1.5dpmag}). In the top right panel, 
 very narrow regions near along the rotational axis are seen to be produced 
 in which the magnetic pressure dominate over the matter pressure 
 (colored by red in the right-hand side). Such regions are formed by turbulent inflows 
of the accreting material from the equator, crossing the butterfly-like regions
 outside the disk, to the polar regions. 
Such flow-in materials obtain sufficient 
 magnetic amplifications when they approach to the rotational axis where
 the differential rotation is stronger, leading to the formations of the 
MHD outflows along the rotational axis (bottom). 

Now we proceed to look more in detail to the properties of
the MHD jets. The jet of
model B10J1.5 has the largest explosion energy with largest
baryon loading. This is because the jet is type II as mentioned.
Since the jet is launched rather earlier ($\sim 1.9 s$) than for
type I, there is much material near the rotational axis, which
makes the baryon load of jets larger for the model. For type I
jets, no systematic dependence of the initial angular momentum
on the masses and the energies is found. We think that this is
because the formation of the type I jets occurs by turbulent
inflows as already mentioned in the previous section. 
The similarity between types I and II is that the jets are at
most subrelativistic (0.07c for model B10J1.5) with the explosion
energy less than $10^{49} \erg$. 
While the ordinary GRBs require the
highly relativistic ejecta, we speculate that thesemildly relativistic
ejecta may be favorable for X-ray flashes (XRF). 

\section{Neutrino-Driven Jet from Collapsars}
\subsection{Numerical Method}
\subsubsection{neutrino pair annihilation in collapsars;
ray-tracing method in special relativity}
 We develop a numerical scheme and 
code for estimating the energy and momentum transfer via 
 neutrino pair annihilation ($\nu + {\bar \nu} \rightarrow e^{-}+ e^{+}$),
 bearing in mind the application to the collapsar models of GRBs \cite{hari09b}.
 To calculate the neutrino flux illuminated from the accretion disk, 
we perform a ray-tracing calculation in the framework of 
special relativity. 

\begin{figure}[tbd]
\includegraphics[width = 80mm]{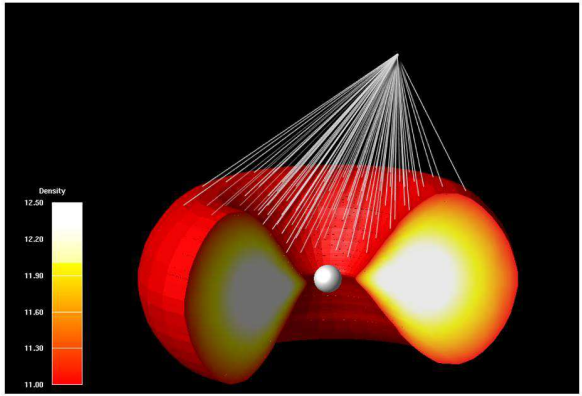}
\caption{An example of the ray-tracing (white line) of neutrinos for estimating the
 neutrino pair annihilation towards a given point outside the accretion disk 
(indicated by red). The central sphere represents the black hole (BH). }
\label{fig:3d_heat}
\end{figure}

Figure \ref{fig:3d_heat} presents
 our calculation concept for the neutrino pair annihilation in collapsars
 (Figure \ref{fig:3d_heat}). For estimating the
 neutrino pair annihilation at a given target, we trace each neutrino trajectory 
(white lines) backwards till it hits to the surface of the accretion disk (colored by 
red). 
We checked the accuracy of our code for various situations, 
and confirmed its validity. 

\subsubsection{two step simulation}
In this section, we show results of the application of newly developed code 
for neutrino heating \cite{hari09b} to collapsars. 
First of all, we note our basic strategy. Our numerical procedure is divided 
into two steps, STEP I and STEP II, depending on the inclusion of neutrino 
heating, since it is numerically expensive. In STEP I, we follow the evolution 
of collapsar from the onset of core-collapse without neutrino heating and 
evaluate the effect of neutrino 
heating in post-processing step. If the neutrino heating meets the condition for 
outflow formation, we go into STEP II in which we restart our simulation 
including neutrino heating from the time when the heating becomes important for 
dynamics. The criterion of this switching is whether $\tau_{\rm dyn}$ is greater 
than $\tau_{\rm heat}$ or not, where $\tau_{\rm dyn}$ and $\tau_{\rm heat}$ is 
the dynamical timescale of materials and the timescale for materials to escape 
from gravitational field by neutrino heating, respectively. Therefore, the 
condition $\tau_{\rm dyn} / \tau_{\rm heat} > 1$ means that materials can escape 
from gravitational field by neutrino energy deposition. Here, the dynamical 
timescale is given by $\tau_{\rm dyn} \equiv \sqrt{3\pi/16G{\bar \rho}}$, where 
$G$ is the gravitational constant and  ${\bar \rho}$ is average density derived 
from spherical mass coordinate $M(r)$ as ${\bar \rho} \equiv 3 M(r) / 4 \pi r^3$. 
Heating time scale is given by $\tau_{\rm heat} \equiv \rho |\Phi_{\rm tot}|/q^+$, 
where $\rho$, $q^+$, and $\Phi_{\rm tot}$ 
is the density, the energy deposition rate by neutrino heating, and the total 
gravitational potential.  

In STEP I, we start our simulation from the onset of core-collapse of the rotating 
massive star. The time-evolution is followed by special relativistic hydrodynamic 
code developed by \cite{taki09,hari09a}. Since the massive star is thought to 
form the BH after the onset  of core-collapse, we simply assume the formation of 
the BH by setting absorptive inner boundary, and put a point mass $M_{\rm BH}$ at the center 
of the progenitor.

In terms of heating, we include the energy deposition and the momentum transfer by 
neutrino-antineutrino pair annihilation for electron type neutrino. To calculate the 
heating rate, we adopt the method in \cite{hari09b}, which is developed for the rapid 
calculation of pair annihilation in collapsars. With this method, we evaluate the heating 
rate in the post-processing step in STEP I. Since the simulation of heating is still 
time-demanding to couple with the hydrodynamics code, we calculate neutrino heating once 
in every 50 hydrodynamical steps in STEP II, which is enough to capture the dynamical 
change of the neutrino flux from the accretion disk. 

We calculated 4 models with various profiles of rotation. 
Model name is given as 'JX' representing the model with 
angular momentum $j=Yj_{\rm lso}$. 

\subsection{Result}
\subsubsection{when the neutrino heating becomes important ?}
\begin{figure}[t]
\includegraphics[width=80mm]{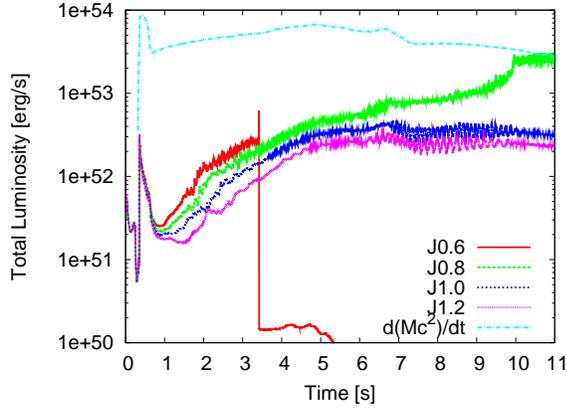}
\caption{Time evolution of total luminosity and the mass accretion rate. Plots for 
the luminosity of J0.6 model (solid), J0.8 model (dashed), J1.0 model (dotted), 
J1.2 model (small dotted), and the mass accretion rate in J0.8 model (dot dashed) 
is shown.}
\label{fig:Ltot}
\end{figure} 
\begin{figure}[t]
\includegraphics[width = 80mm]{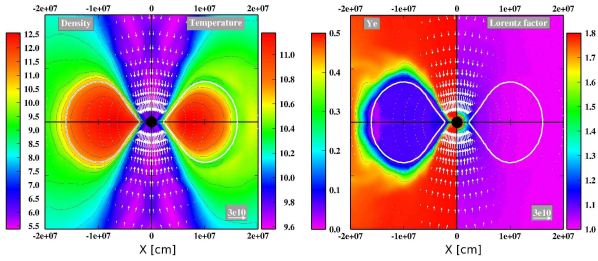}
\caption{Hydrodynamic configuration when the accretion disk is in 
a stationary state (here, 9.1 s after the onset of core-collapse). 
In the left panel, the logarithmic density (in$\gpcmc$, left-half) and
 temperature (in $K$, right-half) are shown. 
The right panel shows the electron fraction (left-half)
 and the Lorentz factor (right-half).}
\label{fig:hydro}
\end{figure} 
\begin{table}[t]
\begin{center}
\begin{tabular}[c]{ccccc}
\hline
Model& Outflow	& $L_{\nu} [{\rm erg \, \, s^{-1}}]$	& $Q_{\nu {\bar \nu}} [{\rm erg \, \, s^{-1}}]$	& ${\dot E_{\rm exp}} [{\rm erg}]	$\\
\hline
J0.6 & No 		& $5.0\times10^{50}$		& $-			   $		& $0.0			 			$\\
J0.8 & Yes 		& $1.2\times10^{53}$		& $3.1\times10^{50}$		& $4.0\times10^{49} 		$\\
J1.0 & No		& $4.1\times10^{52}$		& $2.9\times10^{49}$		& $0.0						$\\
J1.2 & No 		& $2.3\times10^{52}$		& $8.9\times10^{48}$		& $0.0			 			$\\
\hline
\end{tabular}
\caption{Properties of our models. 
Contents of each cell is, whether the neutrino-driven outflows are formed 
(yes or no), the total neutrino luminosity $L_{\nu}$, the neutrino heating rate by 
pair annihilation $Q_{\nu {\bar \nu}}$, and the increasing rate of explosion energy 
${\dot E_{\rm exp}}$. }
\label{Table:outflow}
\end{center}
\end{table}
\begin{figure*}[t]
\includegraphics[width=170mm]{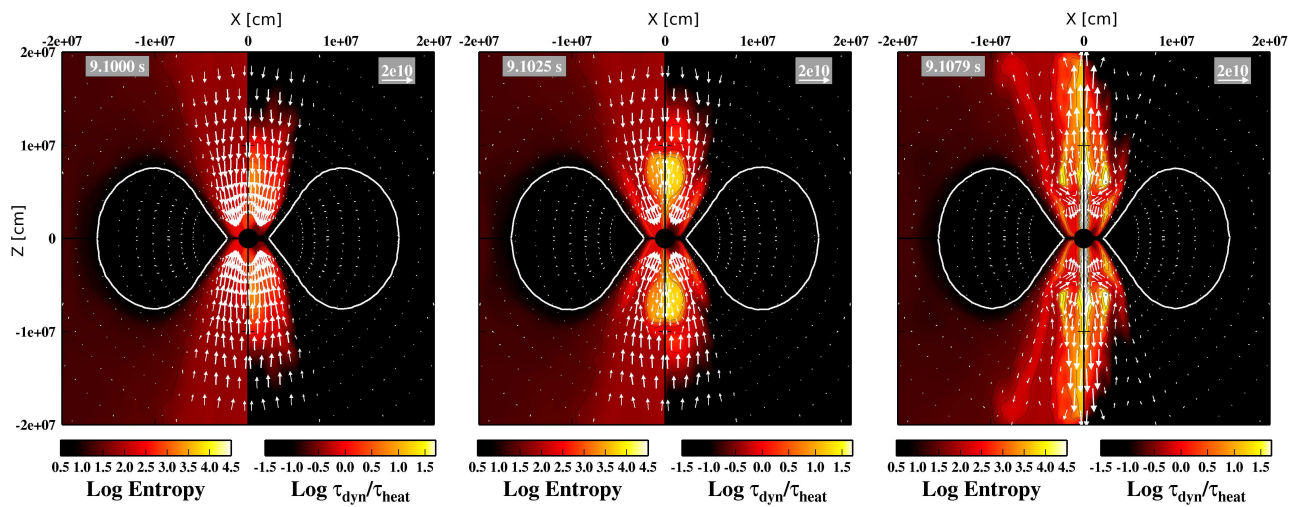}
\caption{Relativistic outflow production in J0.8 model. 
In each panel, logarithmic contour of the entropy (left) and the ratio of the heating 
timescale to the dynamical timescale $\tau_{\rm dyn}/\tau_{\rm heat}$ (right) are 
drawn. Each panel corresponds to 9.1000 s (left), 9.1025 s (center), and 9.1079 s 
(right) after the onset of core-collapse, respectively. The white solid line denotes the area 
where the density is equal to $10^{11} \gpcmc$, showing the surface of the accretion 
disk. }
\label{fig:J0.8O1.2ht}
\end{figure*} 
First of all, we mention about 
the evolution of collapsars without neutrino heating for 11 s (STEP I). All of our 
models initially collapse into the inner radius. When relatively rapidly rotating 
matter with, especially at the edge of Fe core, reaches near the inner radius, 
material crosses the equatorial plane and a shock is formed there ($\approx$ 0.8 s). 
This shock efficiently converts kinetic energy to thermal energy, then thick disk is 
formed in all models. At 2.0 s after the onset of core-collapse, $M_{\rm BH}$ becomes 
greater than 2 $\Ms$ in each model, which would safely verify our assumption of BH 
formation. Figure \ref{fig:hydro} shows the stationary disk at 9.1 s after the onset of 
core-collapse in J0.8 model. We find that large and hot disk can be formed for models 
with angular momentum smaller than $j_{rm lso}$. 

Here after we move to the effect of neutrinos emitted from the accretion disk. 
One of the most important ingredients for neutrino-driven outflow would be the total luminosity
of neutrino, which will be potentially converted to the thermal energy. 
Figure \ref{fig:Ltot} shows time evolution of the neutrino luminosity. 
We fine that after several seconds (namely 5 s), the neutrino luminosity can become greater than 
$10^{53} \ergps$. We also find that the criterion for outflow formation 
$\tau_{\rm dyn} / \tau_{\rm heat} > 1$ is safely satisfied after about 9.0 s only in 
J0.8 model, indicating that outflow would be launched. Therefore we move onto STEP II and 
restart our simulation including neutrino heating from 9.0 s. 

\subsubsection{formation of neutrino-driven outflow}
In STEP II, we find that just after we restart with neutrino heating, an outflow is 
launched from the polar funnel in J0.8 model, while not in other models. Table 
\ref{Table:outflow} shows the main results of this work. In this table, the total 
luminosity $L_{\rm tot}$, the total energy deposition rate $Q_{\nu {\bar \nu}}$ , and 
the luminosity of the explosion energy ${\dot E}_{\rm exp}$ of our models are shown, 
which are given by the average value from 9.1000 s till 9.2044 s after the onset of 
core-collapse. As shown, ${\dot E}_{\rm exp}$ is non-zero ($4.0\times10^{49} \ergps$) 
only for J0.8 model. This outflow is driven by neutrino-antineutrino pair annihilation, 
which we will discuss below. Amang all models, J0.8 model has the largest luminosity 
$L_{\rm tot} \approx 1.2 \times 10^{53} \ergps$ and energy deposition rate 
$Q_{\nu {\bar \nu}} \approx 3.1 \times 10^{50} \ergps$, respectively, 
providing the energy conversion efficiency of 0.26 \%. $Q_{\nu {\bar \nu}}$ is 
comparable to the artificial energy deposition rate in previous studies 
(e.g., \cite{macf99,aloy00}), indicating the analogeous outcomes in the outer layer. 
We note that ${\dot E}_{\rm exp}$ of J0.8 model is consistent with typical energy of 
$\gamma$-ray in the rest frame of GRBs ($\approx 10^{51} \ergps$) if this outflow 
continues for about 100 s and the conversion rate to kinetic energy is several 10 \% 
(e.g., \cite{frail01}), indicating the connection between our model and GRBs. 

\section{Summary}
We have performed a series of numerical simulation of collapsars 
from core-collapsing phase of massive star. 
We focus on indevestigating whether outflow can be launched by MHD and/or neutrino 
process. 
In terms of MHD process, we find that initial weak poloidal magnetic field is much amplified 
inside the disk by compression. Such strong magnetic field can amplify the toroidal magnetic field. 
We find that after several seconds, magnetic pressure dominates over the thermal pressure, 
and trigers the formation of outflow. 
We also find that there are two types of MHD outflows in long-term evolution from core-collapse. 
Although these outflows are weak $(\approx 10^{48} \ergps)$, it can be candidate for the source of 
XRF. 

As for the neutrino process, we developed numerical code for pair neutrino annihilation, 
and applied it to collapsars. We find that the condition for the outflow production 
by neutrino-antineutrino pair annihilation is fulfilled inside 100 km on 
the axis at 9 s after core-collapse. 
Including the energy deposition by neutrino, we find that neutrino-driven 
outflow is formed along the rotational axis. 
In addition, the outflow becomes relativistic. Such relativistic outflow 
can keep collimated till it penetrate the outer layer. 
These features suggests the connection between 
neutrino-driven outflows and GRBs.

\bigskip 

\acknowledgements{S.H. is grateful to T. Kajino for fruitful discussions. 
T.T. and K.K. express thanks to K. Sato and S. Yamada for continuing encouragements.
Numerical computations were in part carried on XT4 and general common use computer 
system at the center for Computational Astrophysics, CfCA, the National Astronomical 
Observatory of Japan.  This study was supported in part by the Grants-in-Aid for the 
Scientific Research from the Ministry of Education, Science and Culture of Japan
(Nos. S19104006, 19540309 and 20740150).

\end{document}